\let\ifarxiv=\iftrue     
\pdfoutput=1

\ifarxiv

\documentclass[12pt,a4paper]{article}
\usepackage[a4paper,text={450pt,650pt},centering]{geometry}

\fi

\ifarxiv\else

\documentclass[11pt,a4paper]{article}
\usepackage{mathptmx}
\usepackage[a4paper,text={130mm,198mm}]{geometry}

\fi

\usepackage{amsfonts}

\usepackage{amsmath,amssymb}
\usepackage[bookmarks=true,hyperfigures=true]{hyperref}
\usepackage{graphicx}
\usepackage[nosort]{cite}

\let\oldbfseries=\bfseries
\let\oldmdseries=\mdseries
\let\oldnormalfont=\normalfont
\renewcommand{\bfseries}{\oldbfseries\boldmath}
\renewcommand{\mdseries}{\oldmdseries\unboldmath}
\renewcommand{\normalfont}{\oldnormalfont\unboldmath}

\allowdisplaybreaks[3]

\numberwithin{equation}{section}

\usepackage[font=small,labelfont=bf,width=0.85\textwidth]{caption}

\providecommand{\hypersetup}[1]{}
\providecommand{\texorpdfstring}[2]{#1}

\hypersetup{plainpages=false}
\hypersetup{pdfpagemode=UseNone}
\hypersetup{bookmarksnumbered=true}
\hypersetup{pdfstartview=FitH}
\hypersetup{colorlinks=false}
\hypersetup{citebordercolor={.5 1 .5}}
\hypersetup{urlbordercolor={.5 1 1}}
\hypersetup{linkbordercolor={1 .7 .7}}

\providecommand{\arxivref}[2]{\href{http://arxiv.org/abs/#1}{#2}}

\providecommand{\href}[2]{#2}
\providecommand{\arxivlink}[1]{\href{http://arxiv.org/abs/#1}{arxiv:#1}}

\newcommand{\be}{\begin{eqnarray}}
\newcommand{\ee}{\end{eqnarray}}
\newcommand{\non}{\nonumber}
\newcommand{\id}{\mathbb{I}}

\newcommand{\A}{\mathop{\cal{A}}\nolimits}

\def\bJ{{\mathbb J}}
\def\bR{{\mathbb R}}
\def\bL{{\mathbb L}}
\def\bQ{{\mathbb Q}}
\def\bC{{\mathbb C}}
\def\bH{{\mathbb H}}

\newcommand{\ihalf}{\sfrac{i}{2}}
\newcommand{\sfrac}[2]{{\textstyle\frac{#1}{#2}}}
\newcommand{\indup}[1]{_{\mathrm{#1}}}
\newcommand{\lrbrk}[1]{\left(#1\right)}

\begin{document}


\thispagestyle{empty}
\phantomsection
\addcontentsline{toc}{section}{Title}

\begin{flushright}\footnotesize%
\texttt{UMTG--269},
\texttt{\arxivlink{1012.3991}}\\
overview article: \texttt{\arxivlink{1012.3982}}%
\vspace{1em}%
\end{flushright}

\begingroup\parindent0pt
\begingroup\bfseries\ifarxiv\Large\else\LARGE\fi
\hypersetup{pdftitle={Review of AdS/CFT Integrability, Chapter III.2: Exact world-sheet S-matrix}}%
Review of AdS/CFT Integrability, Chapter III.2:\\
Exact world-sheet $S$-matrix
\par\endgroup
\vspace{1.5em}
\begingroup\ifarxiv\scshape\else\large\fi%
\hypersetup{pdfauthor={Changrim Ahn and Rafael I. Nepomechie}}%
Changrim Ahn ${}^{1}$ and Rafael I.\ Nepomechie ${}^{2}$
\par\endgroup
\vspace{1em}
\begingroup\itshape
 ${}^{1}$ Department of Physics and Institute for the Early Universe,
       Ewha Womans University,
       Seoul 120-750, South Korea
\vspace{1ex}
      
   ${}^{2}$ Physics Department, P.O. Box 248046, University of Miami,
       Coral Gables, FL 33124 USA
\par\endgroup
\vspace{1em}
\begingroup\ttfamily
ahn@ewha.ac.kr; nepomechie@physics.miami.edu
\par\endgroup
\vspace{1.0em}
\endgroup

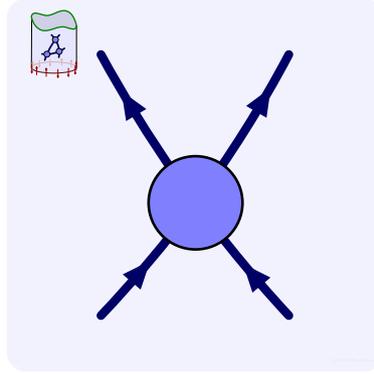
\begin{center}
\includegraphics[width=5cm]{TitleIII2.mps}
\vspace{1.0em}
\end{center}

\paragraph{Abstract:}
We review the derivation of the $S$-matrix for planar ${\cal N}=4$
supersymmetric Yang-Mills theory and type IIB superstring theory on an
$AdS_5\times S^5$ background.  After deriving the $S$-matrix for
the $su(2)$ and $su(3)$ sectors at the one-loop level based on
coordinate Bethe ansatz, we show how $su(2|2)$ symmetry leads to the
exact asymptotic $S$-matrix up to an overall scalar function.  We then
briefly review the spectrum of bound states by relating these states
to simple poles of the $S$-matrix.  Finally, we review the derivation
of the asymptotic Bethe equations, which can be used to determine the
asymptotic multiparticle spectrum.

\ifarxiv\else
\fi
\hypersetup{pdfsubject={MSC (2010): 81T30, 81U15, 16T25, 17B80}}%

\ifarxiv\else
\fi
\hypersetup{pdfkeywords={exact S-matrix, Yang-Baxter equation, Bethe ansatz, AdS/CFT}}%

\newpage


\setcounter{footnote}{0}

\section{Introduction}\label{sec:intro}

$S$-matrices are quantum mechanical
probability amplitudes between incoming and outgoing
on-shell particle states.
Exact factorized $S$-matrices have played a key role in the development of
integrable models \cite{ZZ}.
Indeed, starting from an exact $S$-matrix, it is in principle
possible to compute the asymptotic spectrum, finite-size effects (L\"uscher corrections,
thermodynamic Bethe ansatz), form factors, and
correlation functions non-perturbatively.

As reviewed in many articles in this volume, planar four-dimensional ${\cal N}=4$
supersymmetric Yang-Mills (SYM) theory and its holographic dual, type
IIB superstring theory on $AdS_{5}\times S^{5}$, are believed to be quantum integrable.
The world-sheet and spin-chain $S$-matrix have been derived based on
an $su(2|2)^{2}$ symmetry in \cite{St}-\cite{AFZ} and will be reviewed here.
This $S$-matrix has been confirmed by various checks.
One of these checks is that the all-loop asymptotic
Bethe ansatz equations (BAEs) \cite{BS2}
can be derived from the exact factorized $S$-matrix using either nested Bethe ansatz or
algebraic Bethe ansatz methods \cite{Be1a, Be1b, MM, dL}.
As a warm up, we first review the computation of the
one-loop $S$-matrix in the $su(2)$ and $su(3)$ sectors,
based on a direct coordinate Bethe ansatz,
using integrable spin-chain Hamiltonians whose eigenvalues are
the anomalous dimensions of scalar operators in planar ${\cal N}=4$ SYM.
Using the $S$-matrices, we show how the bound-state spectrum can be constructed.
Finally, we show how imposing periodicity on the asymptotic multiparticle
wavefunction leads to the asymptotic Bethe equations, which can be used
to determine the asymptotic multiparticle spectrum.

The outline of this chapter is as follows.  In Sec. \ref{sec:bulk} we
review the derivation of the exact ${\cal N}=4$ SYM $S$-matrix, first by
coordinate Bethe ansatz for one-loop order, and then by utilizing $su(2|2)$ symmetry
for all-loop order. We also discuss the spectrum of bound states. In
Sec. \ref{sec:ABA} we review the derivation of the asymptotic Bethe
equations, first for the $su(2)$ and $su(3)$ sectors, and then for
the full theory.

\section{Exact \texorpdfstring{$S$}{S}-matrix}\label{sec:bulk}

\subsection{Coordinate Bethe ansatz}\label{coordinate}

For the planar ${\cal N}=4$ SYM theory, we are interested in SYM composite operators,
\be
{\rm Tr}\left[{\cal O}_1{\cal O}_2\cdots{\cal O}_L\right],\qquad
{\cal O}_i \in \{ D^{n}\Phi \,, D^{n} \Psi \,, D^{n} F \} \,,
\label{trace}
\ee
where all operators are at the same spacetime point.
It is useful to associate the composite operators with state vectors of a quantum
spin chain.
The BPS operator ${\rm Tr}[Z^L]$, where $Z$ is one of the scalars 
$\Phi$, is the vacuum state $|0\rangle$.
This choice of vacuum breaks the global $psu(2,2|4)$ symmetry down to 
$su(2|2) \otimes su(2|2)$.
Other composite operators which are obtained by replacing some $Z$'s 
with certain other SYM fields (``impurities'')
are mapped to  excited states over the vacuum:
\be
|\stackrel{\stackrel{1}{\downarrow}}
{Z}\cdots Z
\stackrel{\stackrel{x_1}{\downarrow}}{\chi}Z\cdots Z
\stackrel{\stackrel{x_2}{\downarrow}}{\chi'}Z\cdots Z
\stackrel{\stackrel{x_M}{\downarrow}}{\chi''}Z\cdots
\stackrel{\stackrel{L}{\downarrow}}{Z}\rangle
\equiv{\rm Tr}\left[Z^{x_1-1}\chi 
Z^{x_2-x_1-1}\chi'\cdots\chi''\cdots\right] \,,
\label{state}
\ee
where 
\be
\chi, \chi', \chi'', \ldots \in \{ \Phi_{a{\dot a}}\,, \Psi_{{\dot a}\alpha}\,, 
\bar\Psi_{a {\dot \alpha}}\,, D_{\alpha {\dot \alpha}}Z \} \,, 
\quad a, {\dot a}=1,2,\quad \alpha, {\dot\alpha}=3,4 \,.
\ee 
All other orientations for the operators ${\cal O}_i$ should be 
regarded as multiple excitations $\chi$ coincident at a single 
site.\footnote{For example, $D\Phi$ is a superposition of $\Phi$ and $DZ$. 
More precisely, the excitations are $Z \mapsto DZ$ and $Z \mapsto 
\Phi$; combining these, one obtains $Z \mapsto DZ \mapsto D\Phi$, or 
equivalently $Z \mapsto \Phi \mapsto D\Phi$.}
Due to the cyclic property of the trace, the state (\ref{state})
should be invariant under a uniform translation $x_k\to x_k +1$.
These excitation states belong to a bifundamental representation of
a centrally extended $su(2|2)_L\otimes su(2|2)_R$, which should also be a
symmetry of the $S$-matrix.
The same structure can be discovered on the string world-sheet action in the
light-cone gauge \cite{McLoughlin, Magro}.

For the $S$-matrix, we focus on a particular class of states,
namely asymptotic states, where the distances
between the impurities $\chi,\chi',\ldots$, are very large:
\be
1\ll x_1\ll x_2\ll\cdots\ll x_M\ll L\to\infty.
\ee
The $S$-matrices are defined as amplitudes between two such asymptotic states.

To illustrate this, we derive the two-particle $S$-matrix directly from the
spin chain using coordinate Bethe ansatz.
For simplicity, we will first consider composite operators in the $su(2)$ sector where
the impurities are a complex scalar field $X$.

The one-loop anomalous dimensions of the $su(2)$ sector are given by the
Hamiltonian of the spin-1/2 ferromagnetic $su(2)$-invariant (``XXX'')
Heisenberg quantum spin-chain model \cite{MZ1}
\be
\Gamma = \frac{\lambda}{8\pi^{2}} H\,, \qquad
H = \sum_{l=1}^{L}\left(1 - {\cal P}_{l, l+1}
\right) \,,
\label{Hamiltonian}
\ee
where $\lambda = g^{2}_{YM} N$ is the 't Hooft coupling, and ${\cal P}$ is the
permutation operator on ${\cal C}^{2} \otimes {\cal C}^{2}$.
We also need to impose a periodic boundary condition by identifying $L+1\equiv 1$.

It is obvious that the vacuum state $|0\rangle$ is an
eigenstate of $H$ with zero energy.
Since $[H\,, S^{z}]=0$, the energy eigenstates can be classified 
according to the number of impurities  (``magnons'').
One-particle excited states with momentum $p$ are given
by\footnote{The
invariance of states by a shift of one site (noted earlier) implies
that the total momentum should vanish.  Therefore, a one-particle
state with nonvanishing momentum is not allowed in a strict sense.
The one- or two-particle states which we consider here can be thought
of as part of an infinitely long chain where these particles are
asymptotically separated from other particles.}
\be
|\psi(p)\rangle = \sum_{x=1}^{L} e^{i p x}
|\stackrel{\stackrel{1}{\downarrow}}{Z} \cdots
\stackrel{\stackrel{x}{\downarrow}}{X}
\cdots
\stackrel{\stackrel{L}{\downarrow}}{Z}\rangle.
\label{oneparticle}
\ee
One can easily check that (\ref{oneparticle})
is an eigenstate of $H$ with eigenvalue $E = \epsilon(p)$, where
\be
\epsilon(p) = 4 \sin^{2} (p/2) \,.
\label{epsilon}
\ee

Two-particle eigenstate can be written as
\be
|\psi(p_1,p_2)\rangle &=& A_{XX}(12)|X(p_1)X(p_2)\rangle+A_{XX}(21)|X(p_2)X(p_1)\rangle,
\label{twoXX}\\
|X(p_i)X(p_j)\rangle&=&\sum_{x_{1}<x_{2}}e^{i(p_{i} x_{1} + p_{j} x_{2})}
|\stackrel{\stackrel{1}{\downarrow}}{Z} \cdots
\stackrel{\stackrel{x_{1}}{\downarrow}}{X}
\cdots
\stackrel{\stackrel{x_{2}}{\downarrow}}{X}
\cdots
\stackrel{\stackrel{L}{\downarrow}}{Z}\rangle.
\label{fphiphi}
\ee

Now we impose that these states satisfy
\be
H |\psi\rangle = E(p_1,p_2) |\psi\rangle
\label{EigenvalueProblem}
\ee
and find that
\be
E = \epsilon(p_{1}) + \epsilon(p_{2})  \,,
\label{2particleenergy}
\ee
where $\epsilon(p)$ is given by (\ref{epsilon}).
This leads to the $X-X$ scattering amplitude given by
\be
A_{XX}(21)&=&S(p_{2}\,, p_{1})A_{XX}(12) \,,
\label{AXX}\\
S(p_{2}\,, p_{1})&=& \frac{u_{2} - u_{1} +i}{u_{2} - u_{1} - i}\,,
\label{Sphiphi}
\ee
where $u_{j} = u(p_{j})$ and
\be
u(p) = \frac{1}{2} \cot (p/2) \,.
\label{up}
\ee

We now consider the more complicated case where there are two
different types of complex scalar fields, namely, $X$ and $Y$.
This is the so-called $su(3)$ sector, which is closed only at one loop.
The ($su(3)$-invariant) Hamiltonian is again given by
(\ref{Hamiltonian}), except now ${\cal P}$ is the
permutation operator on ${\cal C}^{3} \otimes {\cal C}^{3}$.
The two-particle eigenstates with one particle of each type are of the form
\be
|\psi\rangle &=&A_{XY}(12)|X(p_1)Y(p_2)\rangle
+A_{XY}(21)|X(p_2)Y(p_1)\rangle \nonumber\\
&+& A_{YX}(12)|Y(p_1)X(p_2)\rangle
+A_{YX}(21)|Y(p_2)X(p_1)\rangle,\label{su3}\\
|\phi_1(p_i)\phi_2(p_j)\rangle&=&
\sum_{x_{1}<x_{2}}
e^{i(p_{i} x_{1} + p_{j} x_{2})}\,
|\stackrel{\stackrel{1}{\downarrow}}{Z} \cdots
\stackrel{\stackrel{x_{1}}{\downarrow}}{\phi_1}
\cdots
\stackrel{\stackrel{x_{2}}{\downarrow}}{\phi_2}
\cdots
\stackrel{\stackrel{L}{\downarrow}}{Z}\rangle.
\label{ansatz}
\ee
Applying the Hamiltonian on $|\psi\rangle$ and imposing the condition (\ref{EigenvalueProblem}),
one finds that the amplitudes should be related by (see e.g. \cite{BV})
\be
\left(
\begin{array}{c}
    A_{XY}(21) \\
    A_{YX}(21)
\end{array} \right) = \left( \begin{array}{cc}
    R(p_{2}\,, p_{1}) & T(p_{2}\,, p_{1}) \\
    T(p_{2}\,, p_{1}) & R(p_{2}\,, p_{1})
    \end{array} \right)
\left(
\begin{array}{c}
    A_{XY}(12) \\
    A_{YX}(12)
    \end{array} \right) \,,
\label{nondiag}
\ee
where the transmission and reflection amplitudes are given by
\be
T(p_{2}\,, p_{1}) = \frac{u_{2}-u_{1}}{u_{2}-u_{1}-i} \,, \qquad
R(p_{2}\,, p_{1}) = \frac{i}{u_{2}-u_{1}-i} \,,
\label{Sphi1phi2}
\ee
respectively.
Combining Eqs.(\ref{Sphiphi}) and (\ref{nondiag}), one can construct
an $su(2)$-invariant
$S$-matrix which connects two amplitudes related by momentum exchange as follows:
\be
\left(
\begin{array}{c}
    A_{XX}(21)\\A_{XY}(21) \\
    A_{YX}(21)\\A_{YY}(21)
\end{array} \right) = \mathbf{S}\cdot
\left(
\begin{array}{c}
    A_{XX}(12)\\A_{YX}(12) \\
    A_{XY}(12)\\A_{YY}(12)
\end{array} \right)
=\left( \begin{array}{cccc}
   S& & & \\
   & T & R& \\
   & R& T& \\
   & & &S
    \end{array} \right)
\left(
\begin{array}{c}
    A_{XX}(12)\\A_{YX}(12) \\
    A_{XY}(12)\\A_{YY}(12)
\end{array} \right) \,.
\label{su2Smatrix}
\ee

At higher loops, the $su(2)$ sector remains closed, but the
Hamiltonian becomes longer ranged. Integrability persists, but only
in a perturbative sense  \cite{Rej}. Correspondingly, one must introduce a perturbative
asymptotic Bethe ansatz, and in particular, an asymptotic $S$-matrix
\cite{St, Su}. That is, in contrast to the
one-loop case (XXX model) where the $S$-matrix is ``local,'' for
higher loops the $S$-matrix is only asymptotic: it applies only to
in-going and out-going particles which are widely separated.

\subsection{Yang-Baxter equation and ZF algebra}

It is not practical to extend the above approach to all loops
and to all sectors of planar ${\cal N}=4$ SYM. Fortunately, there is an
alternative approach -- based on symmetry -- to derive an exact
asymptotic $S$-matrix which is valid for any value of `t Hooft coupling
constant. To this end, it is convenient to introduce
Zamolodchikov-Faddeev (ZF) operators \cite{ZZ, Fa} to define particle states.
Using the ZF operators one can reformulate the derivation of the $S$-matrix into an
algebraic problem.
In Eq.(\ref{ansatz}), we have introduced an asymptotic two-particle state as a superposition of
plane waves.
Now we express these states in terms of creation (ZF) operators acting on the vacuum state
as follows:
\be
|\phi_1(p_i)\phi_2(p_j)\rangle\equiv A_{\phi_1}^{\dagger}(p_i)A_{\phi_2}^{\dagger}(p_j)|0\rangle.
\ee

As can be noticed in (\ref{trace}), the ZF operators corresponding to
the elementary fields of ${\cal N}=4$ SYM can be denoted by
$A_{i{\dot i}}^{\dagger}$,
where the index $i=(a,\alpha)=1,2,3,4$ and similarly for ${\dot i}$.
A very remarkable feature of the AdS/CFT $S$-matrix is that it is
factorized into a tensor product of two identical $S$-matrices, one acting on the index $i$
and the other on ${\dot i}$:
\be
{\mathbb S}=S\otimes{\dot S} \,.
\label{full}
\ee
A natural way to describe the factorized $S$-matrix is to introduce ``quark'' ZF operators
$A_{i}^{\dagger}$ and identify $A_{i{\dot i}}^{\dagger}$ with the tensor product of the
quark ZF operators by
\be
A_{i{\dot i}}^{\dagger}(p)=A_{i}^{\dagger}(p)\otimes A_{\dot i}^{\dagger}(p).
\ee
By the factorization property, it is enough now to consider only $A_{i}^{\dagger}$ sector
for our discussion.

The bulk $S$-matrix elements
$S_{i\, j}^{i' j'}(p_{1}, p_{2})$ define the ZF algebra relation
\be
A_{i}^{\dagger}(p_{1})\, A_{j}^{\dagger}(p_{2}) =
S_{i\, j}^{i' j'}(p_{1}, p_{2})\,
A_{j'}^{\dagger}(p_{2})\, A_{i'}^{\dagger}(p_{1}) \,,
\label{bulkS1}
\ee
where summation over repeated indices is always understood.
It is convenient to arrange these matrix elements into a $16 \times 16$
matrix $S$ as follows,
\be
S = S_{i\, j}^{i' j'} e_{i\, i'}\otimes e_{j\, j'}\,,
\label{bulkS2}
\ee
where $e_{i j}$ is the usual elementary $4 \times 4$ matrix whose
$(i, j)$ matrix element is 1, and all others are zero.

As is well known \cite{ZZ}, starting from $A_{i}^{\dagger}(p_{1})\,
A_{j}^{\dagger}(p_{2})\, A_{k}^{\dagger}(p_{3})$, one can arrive at
linear combinations of $A_{k''}^{\dagger}(p_{3})\,
A_{j''}^{\dagger}(p_{2})\, A_{i''}^{\dagger}(p_{1})$ by applying the
relation (\ref{bulkS1}) three times, in two different ways.  The
consistency condition is the Yang-Baxter equation,
\be
S_{12}(p_{1}, p_{2})\, S_{13}(p_{1}, p_{3})\, S_{23}(p_{2}, p_{3})\ =
S_{23}(p_{2}, p_{3})\, S_{13}(p_{1}, p_{3})\, S_{12}(p_{1}, p_{2}) \,.
\label{YBE}
\ee
We use the standard convention $S_{12} = S \otimes \id$, $S_{23}
= \id \otimes S$, and $S_{13} = {\cal P}_{12}\, S_{23}\, {\cal P}_{12}$,
where ${\cal P}_{12} = {\cal P} \otimes \id$, ${\cal P} = e_{i\, j} \otimes
e_{j\, i}$ is the permutation matrix, and $\id$ is the four-dimensional
identity matrix. The ZF algebra (\ref{bulkS1}) also implies the bulk
unitarity equation
\be
S_{12}(p_{1}, p_{2})\, S_{21}(p_{2}, p_{1}) = \id \,,
\label{bulkunitarity}
\ee
where $S_{21} = {\cal P}_{12}\, S_{12}\, {\cal P}_{12}$.

Solving the Yang-Baxter equation can be complicated.
Fortunately, as we shall see below, $su(2|2)$ symmetry suffices to
determine the AdS/CFT $S$-matrix (in the fundamental representation) 
-- there is no need to solve the Yang-Baxter equation, as it is 
automatically satisfied.

\subsection{Centrally extended \texorpdfstring{$su(2|2)$}{su(2|2)}}

The centrally extended $su(2|2)$ algebra consists of the rotation
generators $\bL_{a}^{\ b}$, $\bR_{\alpha}^{\ \beta}$, the supersymmetry
generators $\bQ_{\alpha}^{\ a}$, $\bQ_{a}^{\dagger \alpha}$, and the
central elements $\bC\,, \bC^{\dagger}\,, \bH$.  \footnote{The central
charge $\bH$ is identified as the world-sheet Hamiltonian.  The
additional central charges $\bC$ and $\bC^{\dagger}$, which are
necessary for having momentum-dependent representations with the
appropriate energy, also appear in the off-shell symmetry algebra of
the gauge-fixed sigma model \cite{Magro}.}
Latin indices $a\,, b\,,
\ldots$ take values $\{1\,, 2\}$, while Greek indices $\alpha\,,
\beta\,, \ldots$ take values $\{3\,, 4\}$. These generators have the
following nontrivial commutation relations \cite{Be1a, Be1b, AFZ}
\be
\left[ \bL_{a}^{\ b}\,, \bJ_{c} \right] &=& \delta_{c}^{b} \bJ_{a} -
\frac{1}{2} \delta_{a}^{b} \bJ_{c}\,, \quad
\left[ \bR_{\alpha}^{\ \beta}\,, \bJ_{\gamma} \right] =
\delta_{\gamma}^{\beta} \bJ_{\alpha} -
\frac{1}{2} \delta_{\alpha}^{\beta} \bJ_{\gamma}\,, \non  \\
\left[ \bL_{a}^{\ b}\,, \bJ^{c} \right] &=& -\delta_{a}^{c} \bJ^{b} +
\frac{1}{2} \delta_{a}^{b} \bJ^{c}\,, \quad
\left[ \bR_{\alpha}^{\ \beta}\,, \bJ^{\gamma} \right] =
-\delta_{\alpha}^{\gamma} \bJ^{\beta} +
\frac{1}{2} \delta_{\alpha}^{\beta} \bJ^{\gamma}\,, \non \\
\Big\{\bQ_{\alpha}^{\ a}\,, \bQ_{\beta}^{\ b}\Big\}&=&
\epsilon_{\alpha \beta}\epsilon^{a b} \bC \,, \quad
\Big\{\bQ_{a}^{\dagger \alpha}\,, \bQ_{b}^{\dagger \beta} \Big\}=
\epsilon^{\alpha \beta}\epsilon_{a b} \bC^{\dagger} \,, \non \\
\Big\{\bQ_{\alpha}^{\ a}\,, \bQ_{b}^{\dagger \beta} \Big\} &=& \delta_{b}^{a}
\bR_{\alpha}^{\ \beta}+ \delta_{\alpha}^{\beta} \bL_{b}^{\ a}
+ \frac{1}{2} \delta_{b}^{a} \delta_{\alpha}^{\beta} \bH \,,
\label{symmetryalgebra}
\ee
where $\bJ_{i}$ ($\bJ^{i}$) denotes any lower (upper) index of a generator,
respectively.

The action of the bosonic generators on the ZF operators is given by
\be
\left[\bL_{a}^{\ b}\,, A_{c}^{\dagger}(p)\right] &=& (\delta_{c}^{b}\delta_{a}^{d} -
\frac{1}{2}\delta_{a}^{b}\delta_{c}^{d}) A_{d}^{\dagger}(p)\,, \quad
\left[\bL_{a}^{\ b}\,, A_{\gamma}^{\dagger}(p)\right] = 0\,, \non \\
\left[\bR_{\alpha}^{\ \beta}\,, A_{\gamma}^{\dagger}(p)\right] &=&
(\delta_{\gamma}^{\beta}\delta_{\alpha}^{\delta} -
\frac{1}{2}\delta_{\alpha}^{\beta}\delta_{\gamma}^{\delta})
A_{\delta}^{\dagger}(p)\,, \quad
\left[\bR_{\alpha}^{\ \beta}\,, A_{c}^{\dagger}(p)\right] = 0\,.
\label{repBulk1}
\ee
The operator relations for supersymmetry generators \footnote{Such
momentum-dependent braiding relations, which are typical for nonlocal
(fractional-spin) integrals of motion, have long been used to
determine $S$-matrices in certain integrable models, see e.g.
\cite{KR, Za, BL}.} 
\be
\bQ_{\alpha}^{\ a}\, A_{b}^{\dagger}(p) &=& e^{- i p/2} \left[
a(p) \delta_{b}^{a} A_{\alpha}^{\dagger}(p) +
A_{b}^{\dagger}(p)\, \bQ_{\alpha}^{\ a} \right] \,, \non \\
\bQ_{\alpha}^{\ a}\, A_{\beta}^{\dagger}(p) &=& e^{- i p/2} \left[
b(p) \epsilon_{\alpha \beta}\epsilon^{a b} A_{b}^{\dagger}(p) -
A_{\beta}^{\dagger}(p)\, \bQ_{\alpha}^{\ a} \right]\,, \non \\
\bQ_{a}^{\dagger \alpha}\, A_{b}^{\dagger}(p) &=& e^{i p/2} \left[
c(p) \epsilon_{a b} \epsilon^{\alpha \beta} A_{\beta}^{\dagger}(p) +
A_{b}^{\dagger}(p)\, \bQ_{a}^{\dagger \alpha} \right] \,, \non \\
\bQ_{a}^{\dagger \alpha}\, A_{\beta}^{\dagger}(p) &=& e^{i p/2} \left[
d(p) \delta_{\beta}^{\alpha} A_{a}^{\dagger}(p) -
A_{\beta}^{\dagger}(p)\, \bQ_{a}^{\dagger \alpha} \right] \,,
\label{repBulk2}
\ee
and the central charges
\be
\bC\, A_{i}^{\dagger}(p) &=& e^{-i p}\left[
a(p) b(p) A_{i}^{\dagger}(p) +
A_{i}^{\dagger}(p)\, \bC \right]\,, \non \\
\bC^{\dagger}\, A_{i}^{\dagger}(p) &=& e^{i p}\left[
c(p) d(p) A_{i}^{\dagger}(p) +
A_{i}^{\dagger}(p)\, \bC^{\dagger} \right] \,, \non \\
\bH\, A_{i}^{\dagger}(p) &=& \left[a(p) d(p) + b(p) c(p)\right] A_{i}^{\dagger}(p) +
A_{i}^{\dagger}(p)\, \bH \,,
\label{repBulk3}
\ee
can be used to act with the generators on multiparticle states.
The ZF operators form a representation of the symmetry algebra
provided $a d - b c = 1$.
The representation is also unitary provided $d= a^{*}\,, c = b^{*}$.
Acting with $\bC$ on both sides of Eq.(\ref{bulkS1}) applied to the 
vacuum state, one can deduce
the further constraint
\be
e^{-ip_1}a(p_1) b(p_1)+e^{-i(p_1+p_2)}a(p_2) b(p_2) = e^{-ip_2}a(p_2) b(p_2)
+e^{-i(p_1+p_2)}a(p_1) b(p_1) \,, \non \\
\ee
which leads to the relation $a(p) b(p) = ig (e^{i p}-1)$, where
$g$ is a constant.  It follows that the parameters can be chosen as
follows \cite{Be1a, AFZ, AF}
\be
a = \sqrt{g}\eta\,, \quad
b = \sqrt{g}\frac{i}{\eta}\left(\frac{x^{+}}{x^{-}}-1\right)\,, \quad
c= -\sqrt{g}\frac{\eta}{x^{+}}\,, \quad
d=\sqrt{g}\frac{x^{+}}{i \eta}\left(1 - \frac{x^{-}}{x^{+}}\right)\,,
\label{BulkParameters}
\ee
where
\be
x^{+}+\frac{1}{x^{+}}-x^{-}-\frac{1}{x^{-}} = \frac{i}{g}\,, \quad
\frac{x^{+}}{x^{-}} = e^{i p}\,, \quad
\eta = e^{i p/4}\sqrt{i(x^{-}-x^{+})} \,.
\label{eta}
\ee
Hence, for a one-particle state,
\be
\bH = -i g \left(x^{+}-\frac{1}{x^{+}}-x^{-}+\frac{1}{x^{-}}\right)
=\sqrt{1+16g^2\sin^2\frac{p}{2}} \,.
\ee
The anomalous dimension $\bH-1$ matches with the weak-coupling result
given by (\ref{Hamiltonian}) and (\ref{epsilon}), provided we make the
identification $g=\sqrt{\lambda}/(4\pi)$. That is, the symmetry
determines the exact dispersion relation, except for the dependence on the
coupling constant. See also \cite{Santambrogio:2002sb}.

The $S$-matrix can be determined (up to a phase) by demanding that it
commute with the symmetry generators. 
That is, starting from
$\bJ\, A_{i}^{\dagger}(p_{1})\, A_{j}^{\dagger}(p_{2})
|0\rangle$ where $\bJ$ is a symmetry generator, and assuming
that $\bJ$ annihilates the vacuum state, one can
arrive at linear combinations of $A_{j'}^{\dagger}(p_{2})\,
A_{i'}^{\dagger}(p_{1}) |0\rangle$ in two different ways, by applying the
ZF relation (\ref{bulkS1}) and the symmetry relations
(\ref{repBulk1}), (\ref{repBulk2})
in different orders.  The consistency condition is
a system of linear equations for the $S$-matrix
elements.  The result for the nonzero matrix elements $S_{i\, j}^{i'
j'}(p_{1}, p_{2})$ is \cite{Be1a,AFZ}
\be
S_{a\, a}^{a\, a} &=& A\,, \quad
S_{\alpha\, \alpha}^{\alpha\, \alpha} = D\,, \non \\
S_{a\, b}^{a\, b} &=& \frac{1}{2}(A-B)\,, \quad
S_{a\, b}^{b\, a} = \frac{1}{2}(A+B) \,, \non \\
S_{\alpha\, \beta}^{\alpha\, \beta} &=& \frac{1}{2}(D-E)\,, \quad
S_{\alpha\, \beta}^{\beta\, \alpha} = \frac{1}{2}(D+E) \,, \non \\
S_{a\, b}^{\alpha\, \beta} &=&
-\frac{1}{2}\epsilon_{a b}\epsilon^{\alpha \beta}\, C \,, \quad
S_{\alpha\, \beta}^{a\, b} =
-\frac{1}{2}\epsilon^{a b}\epsilon_{\alpha \beta}\, F \,, \non \\
S_{a\, \alpha}^{a\, \alpha} &=& G\,, \quad
S_{a\, \alpha}^{\alpha\, a} = H \,, \quad
S_{\alpha\, a}^{a\, \alpha} = K\,, \quad
S_{\alpha\, a}^{\alpha\, a} = L \,,
\label{bulkS3}
\ee
where $a\,, b \in \{1\,, 2\}$ with $a \ne b$;
$\alpha\,, \beta \in \{3\,, 4\}$ with $\alpha \ne \beta$; and
\be
A &=& S_{0}\frac{x^{-}_{2}-x^{+}_{1}}{x^{+}_{2}-x^{-}_{1}}
\frac{\eta_{1}\eta_{2}}{\tilde\eta_{1}\tilde\eta_{2}} \,, \non \\
B &=&-S_{0}\left[\frac{x^{-}_{2}-x^{+}_{1}}{x^{+}_{2}-x^{-}_{1}}+
2\frac{(x^{-}_{1}-x^{+}_{1})(x^{-}_{2}-x^{+}_{2})(x^{-}_{2}+x^{+}_{1})}
{(x^{-}_{1}-x^{+}_{2})(x^{-}_{1}x^{-}_{2}-x^{+}_{1}x^{+}_{2})}\right]
\frac{\eta_{1}\eta_{2}}{\tilde\eta_{1}\tilde\eta_{2}}\,, \non \\
C &=& S_{0}\frac{2i x^{-}_{1} x^{-}_{2}(x^{+}_{1}-x^{+}_{2}) \eta_{1} \eta_{2}}
{x^{+}_{1} x^{+}_{2}(x^{-}_{1}-x^{+}_{2})(1 - x^{-}_{1} x^{-}_{2})}
\,, \qquad
D = -S_{0}\,, \non \\
E &=&S_{0}\left[1-2\frac{(x^{-}_{1}-x^{+}_{1})(x^{-}_{2}-x^{+}_{2})
(x^{-}_{1}+x^{+}_{2})}
{(x^{-}_{1}-x^{+}_{2})(x^{-}_{1} x^{-}_{2}-x^{+}_{1}
x^{+}_{2})}\right]\,, \non \\
F &=& S_{0}\frac{2i(x^{-}_{1}-x^{+}_{1})(x^{-}_{2}-x^{+}_{2})(x^{+}_{1}-x^{+}_{2})}
{(x^{-}_{1}-x^{+}_{2})(1-x^{-}_{1} x^{-}_{2})\tilde\eta_{1} \tilde\eta_{2}}\,,
\non \\
G &=&S_{0}\frac{(x^{-}_{2}-x^{-}_{1})}{(x^{+}_{2}-x^{-}_{1})}
\frac{\eta_{1}}{\tilde\eta_{1}}\,, \qquad
H =S_{0}\frac{(x^{+}_{2}-x^{-}_{2})}{(x^{-}_{1}-x^{+}_{2})}
\frac{\eta_{1}}{\tilde\eta_{2}}\,, \non \\
K &=&S_{0}\frac{(x^{+}_{1}-x^{-}_{1})}{(x^{-}_{1}-x^{+}_{2})}
\frac{\eta_{2}}{\tilde\eta_{1}}\,, \qquad
L = S_{0}\frac{(x^{+}_{1}-x^{+}_{2})}{(x^{-}_{1}-x^{+}_{2})}
\frac{\eta_{2}}{\tilde\eta_{2}}\,,
\label{bulkS4}
\ee
where $x^{\pm}_{i} = x^{\pm}(p_{i})$ and
\be
\eta_{1} = \eta(p_{1}) e^{i p_{2}/2}\,, \quad  \eta_{2}=\eta(p_{2})\,,
\quad  \tilde\eta_{1} =\eta(p_{1})\,, \quad  \tilde\eta_{2}
=\eta(p_{2})e^{i p_{1}/2} \,,
\label{etas}
\ee
where $\eta(p)$ is given in (\ref{eta}).  This $S$-matrix satisfies the
standard Yang-Baxter equation (\ref{YBE}). It also satisfies the
unitarity equation (\ref{bulkunitarity}), provided that the scalar
factor obeys
\be
S_{0}(p_{1}, p_{2})\, S_{0}(p_{2}, p_{1}) = 1 \,.
\label{bulkunitarityscalar}
\ee
In order to determine $S_0$, one should impose on the full $S$-matrix (\ref{full})
crossing symmetry and other physical
requirements, which will be explained in the next chapter of this volume \cite{VieVol}.
The final result is given by
\be
S_{0}(p_{1}, p_{2})^2=\frac{x_1^--x_2^+}{x_1^+-x_2^-}\,\frac{1-\frac{1}{x_1^+ x_2^-}}
{1-\frac{1}{x_1^-x_2^+}}\,\sigma(p_{1}, p_{2})^{2},
\label{dressingphase}
\ee
where the dressing factor $\sigma(p_{1}, p_{2})$ is called
the BES/BHL phase factor \cite{BHL, BES}.

We remark that the above $S$-matrix is in fact in the ``string
frame'' (or ``basis'') \cite{AFZ}.  Starting from the spin chain one
obtains the $S$-matrix instead in the ``spin-chain frame,'' where
(\ref{etas}) is replaced by
\be
\eta_{1} = \eta(p_{1}) \,, \quad  \eta_{2}=\eta(p_{2})\,,
\quad  \tilde\eta_{1} =\eta(p_{1})\,, \quad  \tilde\eta_{2}
=\eta(p_{2}) \,.
\ee
The $S$-matrix in the spin-chain frame satisfies a ``twisted'' version
of the Yang-Baxter equation, rather than (\ref{YBE}).

We also remark that the $su(2|2)$ $S$-matrix is closely related 
\cite{Be1b, MM} to Shastry's $R$-matrix \cite{Shastry1, Shastry2} for the Hubbard model.

\subsection{Bound states}\label{sec:BS}

So far we have considered two-particle asymptotic scattering states.
The two particles carrying real momenta can be widely separated.
Another interesting case occurs when the two particles are closely localized and behave as
a single particle.
This kind of localized state is the bound state \cite{Do1, Do2}.

As a first example, let us consider again the $su(2)$ sector at one loop.
In terms of
\be
x=\frac{x_1+x_2}{2},\quad r=x_2-x_1,\quad p_{1,2}=\frac{p}{2}\pm k,
\ee
we can reexpress the two-particle state (\ref{twoXX}) as
\be
|\psi\rangle &=& \sum_{x,\ r}e^{ipx}\left(A_{XX}(12)e^{-ikr}+A_{XX}(21)e^{ikr}\right)
\stackrel{r}{|Z\cdots\overbrace{XZ\cdots ZX}\cdots{Z}\rangle}.
\ee
Notice that $r>0$ by definition.
To have a localized wave, the amplitude should decay exponentially as the distance $r$ increases.
This can be satisfied if we take $k=iq$ ($q>0$) and $A_{XX}(12)=0$.
From Eq.(\ref{AXX}) this leads to a condition that $S(p_2,p_1)$ should have a pole.
In other words, a simple pole of the $S$-matrix corresponds to a bound state.
In terms of $u$-variables, this condition is satisfied by $u_{2,1}=u\pm i/2$ as one can
see from (\ref{Sphiphi}).
This is an example of a so-called string solution, of size $2$.
Following a similar procedure, one can find that the higher bound-state poles of the
$S$-matrices can be obtained when the particles carry momenta
\be
u^{(n)}_{j}=u+i\ \frac{2j-n-1}{2},\qquad j=1,\ldots,n.
\ee
This is a string of size $n$.
The energy of this particle can be obtained from (\ref{epsilon})
\be
\epsilon_n(u)=\frac{n}{u^2+n^2/4}.
\ee

Now consider the more complicated case of the $su(3)$ sector, for which the two-particle
eigenstates are given by (\ref{su3}) and (\ref{ansatz}).
By the same argument as above, the localized state is possible when $u_2-u_1=i$.
This leads to $A_{XY}(12)=A_{YX}(12)=0$ from (\ref{nondiag}) and $A_{XY}(21)=A_{YX}(21)$ because
the residues of $T$ and $R$ in (\ref{Sphi1phi2}) are the same.
Therefore, the localized state can be written as
\be
|\psi\rangle &\sim& \sum_{x,\ r}e^{ipx}e^{ikr}\left[
\stackrel{r}{|Z\cdots\overbrace{XZ\cdots ZY}\cdots{Z}\rangle}
+\stackrel{r}{|Z\cdots\overbrace{YZ\cdots ZX}\cdots{Z}\rangle}\right],
\ee
where $X$ and $Y$ appear symmetrically.

The bound states for generic value of `t Hooft coupling constant can be constructed
in a similar way.
Combining two factors of the amplitude $A$ (\ref{bulkS4}) with (\ref{dressingphase}),
the $S$-matrix of the $su(2)$ sector (in the spin-chain frame) is given by
\be
S(p_1,p_2)=\frac{x^+_1-x^-_2}{x^-_1-x^+_2}\frac{1-\frac{1}{x_1^+x_2^-}}
{1-\frac{1}{x_1^-x_2^+}} \sigma(p_1,p_2)^{2} \,.
\ee
This amplitude has two simple poles at $x_1^-=x_2^+$ and $x^-_1=1/x_2^+$.
Let us consider first the former case for general higher-order bound states where
simple poles appear
\be
x^-_1=x_2^+,\quad x^-_2=x_3^+,\quad\cdots,\quad x_{n-1}^-=x_{n}^+.
\ee
With these bound-state conditions, one can easily show that the momentum ($p$) and
energy ($\bH$) are given by
\be
\frac{X^{+}}{X^{-}}&=&e^{ip},\qquad X^++\frac{1}{X^+}-X^{-}-\frac{1}{X^-}=\frac{in}{g}
\label{BPS} \\
\bH &=& -i g \left(X^{+}-\frac{1}{X^{+}}-X^{-}+\frac{1}{X^{-}}\right)
=\sqrt{n^2+16g^2\sin^2\frac{p}{2}} \,,
\ee
and satisfy the BPS (shortening) condition in (\ref{BPS}) if we identify
\be
X^{-}\equiv x_n^-,\quad{\rm and}\quad X^{+}\equiv x_1^+.
\ee
The other pole at $x^-_1=1/x_2^+$ cannot satisfy this condition and
leads to non-BPS states.

The situation for the full $su(2|2)$ $S$-matrix is more complicated even though
the locations of poles are the same as in the $su(2)$ sector.
The $M$-particle bound states belong to an atypical totally symmetric representation of
the centrally extended $su(2|2)$ algebra.
This representation has dimension $2M|2M$ and can be realized on the graded vector space
where the basis is given by
\begin{itemize}
\item $M+1$ bosonic states: symmetric in $a_i$: $\vert e_{a_1\cdots a_M}\rangle$, where $a_i=1,2$
are bosonic indices.
\item $M-1$ bosonic states: symmetric in $a_i$: $\vert e_{a_1\cdots a_{M-2}\alpha_1\alpha_2}\rangle$,
where $\alpha_i=3,4$ are fermionic indices.
\item $2M$ fermionic states: symmetric in $a_i$: $\vert e_{a_1\cdots a_{M-1}\alpha}\rangle$,
where $\alpha=3,4$.
\end{itemize}
An efficient realization of this representation is to introduce \cite{AF}
a vector space of analytic
functions of two bosonic variables $w_a$ and two fermionic variables $\theta_{\alpha}$.
For example, the $8$-dimensional states for $M=2$ can be given by
\be
|e_{1}\rangle &=& \frac{w_{1} w_{1}}{\sqrt2} \,, \qquad
|e_{2}\rangle = w_{1} w_{2} \,, \qquad
|e_{3}\rangle = \frac{w_{2} w_{2}}{\sqrt2} \,, \qquad
|e_{4}\rangle = \theta_{3} \theta_{4} \,, \non \\
|e_{5}\rangle &=& w_{1} \theta_{3} \,, \qquad \
|e_{6}\rangle = w_{1} \theta_{4} \,, \qquad \
|e_{7}\rangle = w_{2} \theta_{3} \,, \qquad \
|e_{8}\rangle = w_{2} \theta_{4} \,.
\label{superspacebasis}
\ee
The $su(2|2)$ generators can be
represented by differential operators on this vector space as follows:
\be
\bL_{a}^{\ b} &=&
w_{a}\frac{\partial}{\partial w_{b}}
-\frac{1}{2}\delta^{b}_{a}w_{c}\frac{\partial}{\partial w_{c}} \,,
\qquad \qquad
\bR_{\alpha}^{\ \beta} =
\theta_{\alpha}\frac{\partial}{\partial \theta_{\beta}}
-\frac{1}{2}\delta^{\beta}_{\alpha}\theta_{\gamma}
\frac{\partial}{\partial \theta_{\gamma}} \,, \non \\
\bQ_{\alpha}^{\ a} &=&
a\, \theta_{\alpha}\frac{\partial}{\partial w_{a}}
+ b\, \epsilon^{a b}\epsilon_{\alpha \beta} w_{b}
\frac{\partial}{\partial \theta_{\beta}} \,, \qquad
\bQ_{a}^{\dagger \alpha} =
d\, w_{a} \frac{\partial}{\partial \theta_{\alpha}}
+ c\, \epsilon_{a b}\epsilon^{\alpha \beta}
\theta_{\beta}
\frac{\partial}{\partial w_{b}} \,, \non \\
\bC &=&
a b\, \left(w_{a}\frac{\partial}{\partial w_{a}}+
\theta_{\alpha}\frac{\partial}{\partial \theta_{\alpha}}\right)\,,
\qquad \quad
\bC^{\dagger} =
c d\, \left(w_{a}\frac{\partial}{\partial w_{a}}+
\theta_{\alpha}\frac{\partial}{\partial \theta_{\alpha}}\right)\,, \non \\
\bH &=&
(a d + b c) \left(w_{a}\frac{\partial}{\partial w_{a}}+
\theta_{\alpha}\frac{\partial}{\partial \theta_{\alpha}}\right)\,.
\label{superspacerep}
\ee
From this, it is straightforward to evaluate how the generators act
on the bound states.

In contrast with the case of the fundamental representation reviewed in the previous subsection,
the $su(2|2)$ symmetry is not enough to determine the bound-state $S$-matrix
completely.
A very important observation is that the fundamental bulk $S$-matrix (\ref{bulkS3})
has a remarkable Yangian symmetry $Y(su(2|2))$ \cite{Be2,To} which can be
used to completely determine the two-particle \cite{AF, dL2} and
general $l$-particle bound state bulk $S$-matrices \cite{AdLT}.  It is
fortunate that such a general way of generating higher-dimensional
$S$-matrices has been found, since the fusion procedure
does not seem to work for AdS/CFT $S$-matrices \cite{AF}.

\section{Asymptotic Bethe equations}\label{sec:ABA}

For a system of $N$ {\em free} particles on a ring of length $L$, the
quantized momenta, and therefore the exact spectrum, are trivially
determined.  For particles which are not free but instead have {\em
integrable} interactions, the problem of determining the spectrum is
much more difficult, but nevertheless is still tractable.  Indeed,
if one knows the (asymptotic) $S$-matrix which satisfies the
Yang-Baxter equations, then in principle it is possible to derive a
set of (asymptotic) Bethe equations which determine the (asymptotic)
quantized momenta, and therefore, the (asymptotic) multiparticle spectrum.  These
(asymptotic) Bethe equations are obtained by imposing periodicity on
the (asymptotic) multiparticle wavefunction.  In the AdS/CFT case,
this task is technically difficult due to the matrix structure of the
$S$-matrix and the complicated functional dependence of its matrix
elements.  Before addressing this problem, it is helpful to consider
some simpler examples.

\subsection{The \texorpdfstring{$S$}{S}-matrix is a phase}\label{sec:phase}

As a first warm-up exercise, let us consider the simple case of a
two-body (asymptotic) $S$-matrix which is a phase rather than a
matrix.\footnote{In this case, the Yang-Baxter equations are
trivially satisfied by the $S$-matrix.}
An example is the magnon-magnon $S$-matrix
in the $su(2)$ sector at one loop,
which is given by (\ref{Sphiphi}), (\ref{up}).
The ZF operator $A^{\dagger}(p)$ does not have an
internal index, and satisfies (cf., (\ref{bulkS1}))
\be
A^{\dagger}(p_{1})\, A^{\dagger}(p_{2}) = S(p_{1},p_{2})\,
A^{\dagger}(p_{2})\, A^{\dagger}(p_{1}) \,.
\label{ZFalgebra}
\ee
Integrability of the model implies that the multiparticle wavefunction is of the
Bethe type. That is, the (asymptotic) eigenstates can be expressed as
\be
|\psi \rangle =\sum_{1\le x_{Q_{1}} \ll \ldots \ll x_{Q_{N}\le L}}
\Psi^{(Q)}(x_{1}, \ldots, x_{N})
|\stackrel{\stackrel{1}{\downarrow}}{Z} \cdots
\stackrel{\stackrel{x_{Q_{1}}}{\downarrow}}{X}
\cdots
\stackrel{\stackrel{x_{Q_{N}}}{\downarrow}}{X}
\cdots
\stackrel{\stackrel{L}{\downarrow}}{Z}\rangle \,,
\label{Psi0}
\ee
where the (asymptotic) $N$-particle wavefunction
in the sector $Q=(Q_{1}, \ldots, Q_{N})$ such that
$x_{Q_{1}} \ll \ldots \ll x_{Q_{N}}$ is given by
\be
\Psi^{(Q)}(x_{1}, \ldots, x_{N}) = \sum_{P} A^{P} e^{i p_{P} \cdot
x_{Q}} \,.
\label{PsiI}
\ee
The sum is over all permutations of $P=(P_{1}, \ldots,
P_{N})$, and
$p_{P} \cdot x_{Q} = \sum_{k=1}^{N} p_{P_{k}} x_{Q_{k}}$. Also,
the coordinate-independent amplitudes $A^{P}$ are related to each
other according to
\be
A^{P} \sim A^{\dagger}(p_{P_{1}}) \ldots A^{\dagger}(p_{P_{N}}) \,.
\label{AI}
\ee
For example, for $N=2$, the wavefunction in the sector $x_{1} \ll x_{2}$
is given by
\be
\Psi^{(12)}(x_{1},x_{2}) = A^{12} e^{i(p_{1}x_{1} +p_{2}x_{2})} +
A^{21}e^{i(p_{2}x_{1} +p_{1}x_{2})} \,, \qquad x_{1} \ll  x_{2} \,.
\ee
Since
\be
A^{21} \sim A^{\dagger}(p_{2}) A^{\dagger}(p_{1}) =
S(p_{2},p_{1})\, A^{\dagger}(p_{1}) A^{\dagger}(p_{2}) \sim
S(p_{2},p_{1})\, A^{12} \,,
\ee
we recover the previous results (\ref{twoXX}), (\ref{fphiphi}),
(\ref{AXX}) upon identifying
\be
A_{XX}(12) = A^{12}\,, \qquad A_{XX}(21) = A^{21}\,.
\ee

We consider a system of $N$ widely-separated particles on a ring of
length $L$.
Periodicity of the wavefunction $\Psi(x_{1}, \ldots, x_{N})$ in
(say) the first coordinate,
\be
\Psi(1,x_{2}, \ldots, x_{N}) = \Psi(L+1,x_{2}, \ldots, x_{N}) \,,
\label{PBCI1}
\ee
implies a relationship between the wavefunctions in the sectors 
$x_{1}\ll \ldots \ll x_{N}$ and $x_{2}\ll \ldots \ll x_{N} \ll x_{1}$:
\be
\Psi^{(1 \ldots N)}(1,x_{2}, \ldots, x_{N}) = \Psi^{(2 \ldots N
1)}(L+1,x_{2}, \ldots, x_{N}) \,.
\label{PBCI2}
\ee
According to (\ref{PsiI}), the wavefunctions in these two sectors are 
given by 
\be
\Psi^{(1 \ldots N)}(1,x_{2}, \ldots, x_{N}) &=& A^{1\ldots
N}e^{i(p_{1}+p_{2}x_{2} + \ldots +p_{N}x_{N})} + \ldots \,, \non \\
\Psi^{(2 \ldots N 1)}(L+1,x_{2}, \ldots, x_{N}) &=&  A^{2\ldots
N 1}e^{i(p_{1}L+ p_{1}+ p_{2}x_{2} + \ldots +p_{N}x_{N})} + \ldots \,, \qquad
\label{PsiI2}
\ee
where we have displayed only the terms which depend on the particular 
combination $p_{2}x_{2} + \ldots +p_{N}x_{N}$. In view of the periodicity 
condition (\ref{PBCI2}), the coefficients $A^{1\ldots N}$ 
and $A^{2\ldots N 1}$ in (\ref{PsiI2}) must be related as follows
\be
A^{1\ldots N} = A^{2\ldots N 1} e^{i p_{1} L} \,.
\label{PBCI3}
\ee
There is another relation between the coefficients $A^{1\ldots N} 
$ and $A^{2\ldots N 1}$ which follows from (\ref{AI}). Indeed, it is 
easy to see that
\be
A^{1\ldots N} &\sim&  A^{\dagger}(p_{1}) A^{\dagger}(p_{2}) \ldots A^{\dagger}(p_{N})  \non \\
&=& \prod_{j=2}^{N} S(p_{1}, p_{j}) A^{\dagger}(p_{2}) \ldots
A^{\dagger}(p_{N}) A^{\dagger}(p_{1}) \sim \prod_{j=2}^{N} S(p_{1},p_{j}) A^{2\ldots N 1}
\label{easy}
\,,
\ee
where we have used (\ref{ZFalgebra}) to move $A^{\dagger}(p_{1})$
to the right successively past all the other ZF operators.  The two 
relations  (\ref{PBCI3}) and (\ref{easy}) imply that 
\be
\prod_{j=2}^{N} S(p_{1},p_{j}) = e^{i p_{1} L} \,.
\ee
Examining the terms in the ellipsis in (\ref{PsiI2})
similarly leads to the (asymptotic) Bethe equations for all the momenta,
\be
\prod_{j=1 \atop j\ne k}^{N} S(p_{k},p_{j}) = e^{i p_{k} L}
\,, \qquad k = 1, \ldots, N \,.
\ee
For a ``local'' $S$-matrix such as the one for the spin-1/2
ferromagnetic Heisenberg chain, these equations are exact for finite
$L$; at least in principle one can solve these equations for the
momenta and therefore compute the exact finite-$L$ spectrum,
\be
{\bf P} = \sum_{k=1}^{N}p_{k}\,, \qquad {\bf E} =  \sum_{k=1}^{N}
\epsilon(p_{k}) \,,
\ee
where $\epsilon(p)$ is the one-particle dispersion relation
(see, e.g. (\ref{epsilon})).  For an asymptotic $S$-matrix such as the one for AdS/CFT, the
asymptotic Bethe equations can be used to determine the spectrum only
asymptotically.  \footnote{Nevertheless,
it is possible to obtain at least a part of the exact spectrum by
other means \cite{Janik}.}

\subsection{The \texorpdfstring{$S$}{S}-matrix is a \texorpdfstring{$4\times 4$}{4x4} matrix}\label{sec:mat}

As a second warm-up exercise, we consider a solution of the
Yang-Baxter equations which is a $4\times 4$ matrix. For simplicity, we further
restrict the $S$-matrix to be $su(2)$-invariant. Hence, we take
\be
S_{j k}^{j' k'}(p_{1},p_{2}) = \frac{1}{u_{1}-u_{2}-i}\left[
(u_{1}-u_{2}) \delta_{j}^{j'}\delta_{k}^{k'}
+ i \delta_{j}^{k'}\delta_{k}^{j'} \right] \,,
\label{Smatrix}
\ee
where again $u_{j}=u(p_{j})$ and $u(p)$ is given by (\ref{up}).
This is in fact the magnon-magnon $S$-matrix in the $su(3)$ sector
which we discussed earlier (\ref{su2Smatrix}).
The ZF operator now has an internal index which can take the values 1 and
2, and satisfies (\ref{bulkS1}). As we shall see, the analysis is 
similar to the one in Sec. \ref{sec:phase}. The new feature is the 
internal symmetry, which is handled neatly by introducing the 
transfer matrix (\ref{transfer}).

The (asymptotic) eigenstates can now be expressed as
\be
|\psi \rangle =\sum_{1\le x_{Q_{1}} \ll \ldots \ll x_{Q_{N}\le L}}
\sum_{i_{1}, \ldots, i_{N}=1}^{2}
\Psi^{(Q)}_{i_{1}\ldots i_{N}}(x_{1}, \ldots, x_{N})
|\stackrel{\stackrel{1}{\downarrow}}{Z} \cdots
\stackrel{\stackrel{x_{Q_{1}}}{\downarrow}}{\phi_{i_{1}}}
\cdots
\stackrel{\stackrel{x_{Q_{N}}}{\downarrow}}{\phi_{i_{N}}}
\cdots
\stackrel{\stackrel{L}{\downarrow}}{Z}\rangle \,,
\ee
where the (asymptotic) $N$-particle wavefunction in the sector $Q=(Q_{1}, \ldots,
Q_{N})$ is given by \footnote{The original papers include 
\cite{Yang1}-\cite{Yang4}. Here
we follow the appendix in \cite{AF2}.}
\be
\Psi^{(Q)}_{i_{1}\ldots i_{N}}(x_{1}, \ldots, x_{N}) = \sum_{P}
A^{P|Q}_{i_{1}\ldots i_{N}} e^{i p_{P} \cdot x_{Q}}
\ee
and
\be
A^{P|Q}_{i_{1}\ldots i_{N}} \sim A^{\dagger}_{i_{Q_{1}}}(p_{P_{1}}) \ldots
A^{\dagger}_{i_{Q_{N}}}(p_{P_{N}}) \,,
\label{APQ}
\ee
cf. (\ref{Psi0})-(\ref{AI}). For $N=2$ in the sector $x_{1}\ll
x_{2}$, upon identifying
\be
A_{\phi_{i} \phi_{j}}(12) = A_{i j}^{12|12}\,, \qquad
A_{\phi_{i} \phi_{j}}(21) = A_{i j}^{21|12}
\ee
where $\phi_{1}=X, \phi_{2}=Y$,
we recover the previous results (\ref{su3})-(\ref{su2Smatrix}). \footnote{For example,
\begin{eqnarray*}
A_{XY}(21) &=& A_{12}^{21|12} \sim A_{1}^{\dagger}(p_{2})
A_{2}^{\dagger}(p_{1}) = S_{12}^{12} A_{2}^{\dagger}(p_{1})A_{1}^{\dagger}(p_{2})
+ S_{12}^{21} A_{1}^{\dagger}(p_{1})A_{2}^{\dagger}(p_{2}) \non \\
&\sim& S_{12}^{12} A_{21}^{12|12}
+ S_{12}^{21} A_{12}^{12|12} = T A_{YX}(12) + R A_{XY}(12) \,, \non
\end{eqnarray*}
which is in agreement with (\ref{nondiag}).  Here the arguments $(p_{2},p_{1})$ of
all the $S$-matrix elements have been suppressed for brevity.}

Proceeding as before, we see that the periodicity of the wavefunction
in the first coordinate,
\be
\Psi_{i_{1}\ldots i_{N}}(1,x_{2}, \ldots, x_{N}) =
\Psi_{i_{1}\ldots i_{N}}(L+1,x_{2}, \ldots, x_{N})
\ee
implies a relationship between the wavefunctions in the sectors 
$x_{1}\ll \ldots \ll x_{N}$ and $x_{2}\ll \ldots \ll x_{N} \ll x_{1}$:
\be
\Psi_{i_{1}\ldots i_{N}}^{(1\ldots N)}(1,x_{2}, \ldots, x_{N})
= \Psi_{i_{1}\ldots i_{N}}^{(2\ldots N 1)}(L+1,x_{2}, \ldots, x_{N}) 
\,.
\ee
This leads to the following relationship between coefficients
\be
A^{1\ldots N|1\ldots N}_{i_{1}\ldots i_{N}}
= A^{2\ldots N 1|2\ldots N 1}_{i_{1}\ldots i_{N}} e^{i p_{1} L} \,.
\label{PBCII1}
\ee
We now proceed to generate from (\ref{APQ}) another relation between 
these two coefficients. Using (\ref{bulkS1}) to
move $A^{\dagger}_{i_{1}}(p_{1})$ to the right successively past all the other ZF
operators, we obtain
\be
A^{1\ldots N|1\ldots N}_{i_{1}\ldots i_{N}}&\sim&
A^{\dagger}_{i_{1}}(p_{1}) A^{\dagger}_{i_{2}}(p_{2}) \ldots A^{\dagger}_{i_{N}}(p_{N})
\non \\
&=&
S_{i_{1} i_{2}}^{a_{2} i'_{2}}(p_{1},p_{2})
S_{a_{2} i_{3}}^{a_{3} i'_{3}}(p_{1},p_{3})
\ldots
S_{a_{N-1} i_{N}}^{i'_{1} i'_{N}}(p_{1},p_{N})
A^{\dagger}_{i'_{2}}(p_{2}) \ldots
A^{\dagger}_{i'_{N}}(p_{N})
A^{\dagger}_{i'_{1}}(p_{1}) \non \\
&\sim&
S_{i_{1} i_{2}}^{a_{2} i'_{2}}(p_{1},p_{2})
S_{a_{2} i_{3}}^{a_{3} i'_{3}}(p_{1},p_{3})
\ldots
S_{a_{N-1} i_{N}}^{i'_{1} i'_{N}}(p_{1},p_{N})
A^{2\ldots N 1|2\ldots N 1}_{i'_{1}\ldots i'_{N}}
 \,.
\label{observe}
\ee
It is very convenient to introduce the so-called (inhomogeneous) transfer matrix
\be
t_{i_{1}\ldots i_{N}}^{i'_{1}\ldots i'_{N}}(p; p_{1},\ldots,p_{N})
\equiv
S_{a_{N} i_{1}}^{a_{1} i'_{1} }(p,p_{1})
S_{a_{1} i_{2}}^{a_{2} i'_{2}}(p,p_{2})
\ldots
S_{a_{N-1} i_{N}}^{a_{N} i'_{N}}(p,p_{N}) \,.
\label{transfer}
\ee
Its value at $p=p_{1}$ is proportional to the coefficient
of $A^{2\ldots N 1|2\ldots N 1}_{i'_{1}\ldots i'_{N}}$ in
(\ref{observe}),
\be
t_{i_{1}\ldots i_{N}}^{i'_{1}\ldots i'_{N}}(p_{1}; p_{1},\ldots,p_{N}) =
-S_{i_{1} i_{2}}^{a_{2} i'_{2}}(p_{1},p_{2})
S_{a_{2} i_{3}}^{a_{3} i'_{3}}(p_{1},p_{3})
\ldots
S_{a_{N-1} i_{N}}^{i'_{1} i'_{N}}(p_{1},p_{N})  \,,
\label{eval}
\ee
since $S_{i j}^{i' j'}(p,p) = -\delta_{i}^{j'}\delta_{j}^{i'}$, as 
one can see from (\ref{Smatrix}).

We demand that $A^{2\ldots N 1|2\ldots N 1}_{i'_{1}\ldots i'_{N}}$
be an eigenvector of the transfer matrix, \footnote{This is necessary
in order to be able to satisfy (\ref{PBCII1}). We note that the transfer
matrix has the commutativity property
\[
\left[t(p; p_{1},\ldots,p_{N}) \,, t(p'; p_{1},\ldots,p_{N}) \right]
= 0 \non
\]
by virtue of the fact that the $S$-matrix satisfies the Yang-Baxter
equation. (See, eg. \cite{QISM1}-\cite{QISM3}.) Hence, the corresponding eigenvectors do not
depend on the value of $p$.}
\be
t_{i_{1}\ldots i_{N}}^{i'_{1}\ldots i'_{N}}(p; p_{1},\ldots,p_{N})\,
A^{2\ldots N 1|2\ldots N 1}_{i'_{1}\ldots i'_{N}} =
\Lambda(p; p_{1},\ldots,p_{N}) A^{2\ldots N 1|2\ldots N 
1}_{i_{1}\ldots i_{N}} \,,
\label{eigen}
\ee
where $\Lambda(p; p_{1},\ldots,p_{N})$ is the corresponding eigenvalue.
It follows from Eqs. (\ref{PBCII1}), (\ref{observe}), 
(\ref{eval}), (\ref{eigen}) that
\be
\Lambda(p_{1}; p_{1},\ldots,p_{N})   = - e^{i p_{1} L} \,;
\ee
and more generally
\be
\Lambda(p_{k}; p_{1},\ldots,p_{N})    = -e^{i p_{k} L} \,, \qquad k = 1, \ldots, N \,.
\label{PBCII2}
\ee

To summarize so far: imposing periodic boundary conditions on the
multiparticle wavefunction has led to the important relations (\ref{PBCII2}).
However, in order to obtain more explicit equations for the momenta, we need the eigenvalues
$\Lambda(p; p_{1},\ldots,p_{N})$ of the transfer matrix (\ref{transfer}).
For the case of the $S$-matrix (\ref{Smatrix}),
the result is well known \cite{QISM1}-\cite{QISM3},
\be
\Lambda(p; p_{1},\ldots,p_{N}) &=&
\frac{1}{\prod_{l=1}^{N}(u-u_{l}-i)}\Bigg\{
\prod_{l=1}^{N}(u-u_{l}+i)
\prod_{l=1}^{m}\left(\frac{u-\lambda_{l}-\frac{i}{2}}{u-\lambda_{l}+\frac{i}{2}}\right) \non\\
& & +
\prod_{l=1}^{N}(u-u_{l})
\prod_{l=1}^{m}\left(\frac{u-\lambda_{l}+\frac{3i}{2}}
{u-\lambda_{l}+\frac{i}{2}}\right) \Bigg\} \,,
\label{wellknown}
\ee
where the ``auxiliary'' Bethe roots $\lambda_{1}, \ldots,
\lambda_{m}$ satisfy the Bethe ansatz equations
\be
\prod_{l=1}^{N}\frac{\lambda_{k}-u_{l}+\frac{i}{2}}{\lambda_{k}-u_{l}-\frac{i}{2}}
=\prod_{j=1 \atop j\ne
k}^{m}\frac{\lambda_{k}-\lambda_{j}+i}{\lambda_{k}-\lambda_{j}-i}
\,, \qquad k = 1, \ldots, m \,.
\label{BAEII}
\ee
Finally, substituting the result (\ref{wellknown}) into (\ref{PBCII2}), we
obtain
\be
\prod_{l=1}^{N}\frac{u_{k}-u_{l}+i}{u_{k}-u_{l}-i}
\prod_{l=1}^{m}\frac{u_{k}-\lambda_{l}-\frac{i}{2}}{u_{k}-\lambda_{l}+\frac{i}{2}}
= -e^{i p_{k} L} \,, \qquad k = 1, \ldots, N \,.
\label{PBCII3}
\ee
The coupled set of equations (\ref{BAEII}) and (\ref{PBCII3}) are the
sought-after (asymptotic) Bethe equations for a system of $N$ particles
on a ring of length $L$ with the two-particle (asymptotic) $S$-matrix (\ref{Smatrix}).

\subsection{AdS/CFT}

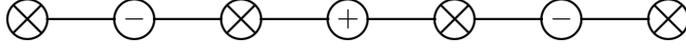
\begin{figure}\centering
\begin{minipage}{260pt}
\setlength{\unitlength}{1pt}%
\small\thicklines%
\begin{picture}(260,20)(-10,-10)
\put(  0,00){\circle{15}}%
\put(  7,00){\line(1,0){26}}%
\put( 40,00){\circle{15}}%
\put( 47,00){\line(1,0){26}}%
\put( 80,00){\circle{15}}%
\put( 87,00){\line(1,0){26}}%
\put(120,00){\circle{15}}%
\put(127,00){\line(1,0){26}}%
\put(160,00){\circle{15}}%
\put(167,00){\line(1,0){26}}%
\put(200,00){\circle{15}}%
\put(207,00){\line(1,0){26}}%
\put(240,00){\circle{15}}%
\put( -5,-5){\line(1, 1){10}}%
\put( -5, 5){\line(1,-1){10}}%
\put( 75,-5){\line(1, 1){10}}%
\put( 75, 5){\line(1,-1){10}}%
\put(155,-5){\line(1, 1){10}}%
\put(155, 5){\line(1,-1){10}}%
\put(235,-5){\line(1, 1){10}}%
\put(235, 5){\line(1,-1){10}}%
\put( 40,00){\makebox(0,0){$-$}}%
\put(120,00){\makebox(0,0){$+$}}%
\put(200,00){\makebox(0,0){$-$}}%
\end{picture}
\end{minipage}

\caption{Dynkin diagram of $su(2,2|4)$.}
\label{fig:DynkinHigher}
\end{figure}

We are finally ready to address the AdS/CFT case, albeit only
sketchily.  The arguments of
Sec.  \ref{sec:mat} leading to (\ref{PBCII2}) carry through
essentially unchanged.\footnote{It is convenient to work in a graded
formalism, where certain minus signs appear.  \cite{MM}} The difficult
step is determining the eigenvalues of the transfer matrix.  Whereas
for the $4 \times 4$ $S$-matrix (\ref{Smatrix}) the result
(\ref{wellknown}) is easily obtained by algebraic Bethe ansatz, for
the larger AdS/CFT $S$-matrix (\ref{bulkS3}),(\ref{bulkS4})  a more general procedure (namely, {\em
nested} algebraic Bethe ansatz) is required \cite{MM}. Alternatively,
the result can be obtained by nested coordinate Bethe ansatz
\cite{Be1a, dL}
or by analytic Bethe ansatz \cite{Be1b}. In this way, one can derive
the $AdS_5/CFT_4$ asymptotic Bethe equations which were first conjectured in
\cite{BS2}. In terms of the compact notation introduced in \cite{BR}, these
equations are given by
\be
U_0=1,\qquad
U_j(x_{j,k})
\mathop{\prod_{j'=1}^7\prod_{k'=1}^{K_{j'}}}_{(j',k')\neq(j,k)}
\frac{u_{j,k}-u_{j',k'}+\ihalf M_{j,j'}}{u_{j,k}-u_{j',k'}-\ihalf M_{j,j'}}
=1, \quad j = 1, \ldots, 7, 
\ee
where $u_{j,k}=g(x_{j,k}+1/x_{j,k})$,
$u_{j,k}\pm i/2=g(x_{j,k}^{\pm}+1/x_{j,k}^{\pm})$, and
$M_{j,j'}$ is the Cartan matrix specified by
Figure \ref{fig:DynkinHigher}. Explicitly,
\be
M=\left(\begin{array}{rrrrrrr}
 &1 \\
1 &-2 &1\\
& 1 & &-1 \\
& & -1 &2 &-1\\
& & & -1 &  & 1\\
& & & & 1 & -2 & 1\\
& & & & & 1 
\end{array} \right) \,,
\ee 
where matrix elements which are zero are left empty.
Also,
\be
U_0=
\prod_{k=1}^{K_4}
\frac{x^+_{4,k}}{x^-_{4,k}}\,,
\quad
U_{2}=U_{6}=1\,, 
\quad
U_1(x)=U_3^{-1}(x)=U_5^{-1}(x)=U_7(x)=
\prod_{k=1}^{K_4}
S\indup{aux}(x_{4,k},x) \non \\
\ee
and
\begin{equation}
U_4(x)=
U\indup{s}(x)
\lrbrk{\frac{x^-}{x^+}}^L
\prod_{k=1}^{K_1}
S^{-1}\indup{aux}(x,x_{1,k})
\prod_{k=1}^{K_3}
S\indup{aux}(x,x_{3,k})
\prod_{k=1}^{K_5}
S\indup{aux}(x,x_{5,k})
\prod_{k=1}^{K_7}
S^{-1}\indup{aux}(x,x_{7,k}).
\end{equation}
Moreover,
\be
S\indup{aux}(x_1,x_2)=
\frac{1-1/x^+_1x_2}{1-1/x^-_1x_2}\,, \qquad
U\indup{s}(x) = \prod_{k=1}^{K_4} \sigma(x,x_{4,k})^{2} \,,
\ee
where $\sigma$ is the dressing phase \cite{BES, VieVol}. The 
anomalous dimensions of a state is given by
\be
\Gamma = 2 i g \sum_{k=1}^{K_{4}}\left(\frac{1}{x^{+}_{4,k}} - 
\frac{1}{x^{-}_{4,k}} \right) \,.
\ee 
For further important details such as the restrictions on the
excitation numbers $K_{1}, \ldots, K_{7}$, the so-called dynamical
transformations relating roots of type 1 and type 3 (and similarly,
roots of type 5 and type 7), and the weak-coupling limit, the reader
should consult \cite{BS2, BR}.

Similarly, starting from the $AdS_4/CFT_3$ $S$-matrix
\cite{AhnNep}, one can derive the corresponding asymptotic Bethe
equations which were first conjectured in \cite{GV}.

\section{Concluding Remarks}\label{sec:Conc}

The all-loop $AdS_{5}/CFT_{4}$ $S$-matrix has further important
applications. In particular,
it is used for computing wrapping corrections via the L\"uscher formula (reviewed
in \cite{Janik}) and finite-size effects via thermodynamic Bethe ansatz
(reviewed in \cite{Bajnok}). A certain Drinfeld twist of this 
$S$-matrix, together with $c$-number diagonal twists of the boundary 
conditions, lead \cite{ABBN} to the deformed Bethe equations of 
Beisert and Roiban \cite{BR, Zoubos}.

The $su(2|2)$ $S$-matrix of $AdS_{5}/CFT_{4}$ also
plays an important role in determining the $S$-matrix of
$AdS_{4}/CFT_{3}$ \cite{AhnNep} (see also \cite{Klose}).
Indeed, the scattering matrices for the two types of
particles (``solitons'' and ``antisolitons'') again have the same $su(2|2)$ matrix structure;
the main difference with respect to the $AdS_{5}/CFT_{4}$ case
is in the scalar factors, which satisfy
new crossing relations. As already noted, this $S$-matrix leads to
the all-loop BAEs conjectured in \cite{GV}.

\section*{Acknowledgments}
We thank N.\ Beisert for his helpful comments.
This work was supported in part by KRF-2007-313-C00150
and WCU grant R32-2008-000-10130-0 (CA),
and by the National Science Foundation under Grants PHY-0554821 and
PHY-0854366 (RN).

\phantomsection

\end{document}